\documentclass[epj]{svjour}

\usepackage{graphicx}
\usepackage{fancyhdr}
\usepackage{amssymb}
\def\makeheadbox{{
 }}
\def\mytitle{My title} 
\def\myauthors{My name}  
\def\mytype{My type of session}
\def\mysession{My session}

\def\mytitle{Di-photon Higgs Decay in SUSY with CP Violation} 
\def\myauthors{S.~Hesselbach, S.~Moretti, S.~Munir, P.~Poulose}    
\def\mytype{Contributed Talk}    
\def\mysession{Colliders - Higgs Phenomenology}

\begin{document}
\title{Di-Photon Higgs Decay in SUSY with CP Violation}
\author{S. Hesselbach%
\inst{1}%
\thanks{\emph{Email:} s.hesselbach@phys.soton.ac.uk}%
 \and
 S. Moretti%
\inst{1,2}%
\thanks{\emph{Email:} stefano@phys.soton.ac.uk}%
 \and
 S. Munir%
\inst{1}%
\thanks{\emph{Email:} shobig@phys.soton.ac.uk}%
 \and
 P. Poulose%
\inst{1,3}%
\thanks{\emph{Email:} poulose@phys.soton.ac.uk}%
}                     
%
%
\institute{School of Physics \& Astronomy, University of Southampton,
  Highfield, Southampton SO17~1BJ, UK
\and Laboratoire de Physique Th\'eorique,
Universit\'e Paris--Sud, F--91405 Orsay Cedex, France
\and Physics Department, IIT Guwahati, Assam, INDIA - 781039}
%
\date{}
\abstract{
Physical Higgs particles in the
Minimal Supersymmetric Standard Model (MSSM) with explicit CP
violation are CP mixed states.
The decay of these Higgs particles can be analysed to study the
CP properties of the MSSM. In the present work we consider the di-photon
channel of the lightest neutral Higgs boson for this purpose. Compared to
earlier studies on effects of scalar/pseudo-scalar mixing, our analysis
also investigates the effect due to Higgs-sfermion-sfermion couplings
along with that of mixing. We find that
a light stop may have a strong impact on the width and Branching
Ratio (BR) of the decay process $H_1\rightarrow \gamma\gamma$, whereas
other light sparticles have only little influence.
In some regions of the MSSM parameter space with large CP-violating
phase $\phi_\mu\sim 90^\circ$
a light ($\sim 200$ GeV) stop can change the di-photon BR by more than 50\%
compared to the case with heavy ($\sim 1 $ TeV) stop and otherwise same
MSSM parameters.
} 
\maketitle
%

\section{Introduction}
\label{intro}
\sloppypar
One of the main objectives of the upcoming Large Hadron Collider (LHC)
is to investigate ElectroWeak Symmetry Breaking (EWSB). While the 
Standard (SM) Higgs mechanism predicts one physical scalar particle, many 
models beyond it have more than one scalar states 
as well as pseudo-scalar 
(and charged) ones. In the presence of CP-violation 
scalar/pseudo-scalar mixing occurs to give the physical Higgs states.
Studying CP properties of the Higgs particles thus becomes an important
feature in distinguishing different models.
Among the new-physics scenarios Supersymmetry (SUSY) is one of the favourites
of particle physicists. LHC will investigate various aspects of SUSY with 
special attention to the Minimal Supersymmetric Standard Model (MSSM).
While the phenomenology of the CP-conserved MSSM 
has been thoroughly studied, many issues of the MSSM with CP violation
are yet to be investigated. 
Many parameters of the MSSM
can well be complex and thus explicitly break CP invariance inducing
CP violation also in the Higgs sector beyond Born approximation
\cite{Pilaftsis:1998dd}.
After elimination of unphysical phases and imposing universality conditions at
the unification scale two independent phases remain,
the phase $\phi_\mu$ of the higgsino mass term $\mu$ and
a common phase $\phi_{A_f}$ of the soft trilinear Yukawa couplings $A_f$
in the sfermion sector \cite{Pilaftsis:1999qt}.
Experimental searches of Electric Dipole Moments (EDMs) 
of electron and neutron put constraints on the CP phases of any model.
In the MSSM with CP violation these constraints can be avoided by taking
the sfermions belonging to the first two generations to be very heavy
(see \cite{Olive:2005ru} for a review).

We consider the di-photon decay mode, $H_1\rightarrow \gamma\gamma$,
of the lightest neutral Higgs boson $H_1$, which involves
direct, i.e.\ leading, effects of the SUSY phases through couplings of
the $H_1$ to SUSY particles in the loops
as well as indirect, i.e.\ sub-leading, effects through the
scalar/pseudo-scalar mixing yielding the Higgs mass-eigenstate $H_1$.
In scenarios with heavy SUSY particles, where  CP violation enters
solely through the scalar/pseudo-scalar mixing, the SUSY CP phases can
result in a strong suppression of the BR of the
decay $H_1\rightarrow \gamma\gamma$ as well as of the rate of the
combined production and decay
process $gg \to H_1 \to \gamma\gamma$ \cite{Choi:2001iu}.
Here, we summarize the results of \cite{Moretti:2007th,Hesselbach:2007en} 
focusing especially on the effects of light SUSY particles
in the decay $H_1\rightarrow \gamma\gamma$.
The analysis of the full production and decay process at the LHC is
postponed to a forthcoming publication \cite{fullprocess}.

\section{\boldmath The $H_1\rightarrow \gamma\gamma$ decay}

As mentioned in the Introduction we consider explicit CP violation
(and assume that the Higgs
vacuum expectation values are real). Thus, in this particular scenario
under study with
common phases for the trilinear couplings and separately for the gaugino 
masses,
we are left with two independent phases after symmetry considerations. As 
intimated, we take these to be $\phi_\mu$ 
and $\phi_{A_f}$.
The leading terms in the CP-violating scalar/pseudo-scalar mixing in
the Higgs sector are proportional to $\mathrm{Im}(\mu A_f)$, hence we
assume $\phi_{A_f} = 0$ and analyse the effects of a non-zero $\phi_\mu$
in the following.
With the help of the publicly available \textsc{Fortran} code \textsc{CPSuperH}
\cite{Lee:2003nta}, version 2, which calculates the mass spectrum and
decay widths of all Higgs bosons along with their couplings to SM and
SUSY particles, we analyse the Higgs decay into the di-photon channel 
in the CP-violating MSSM and compare it with that of the CP-conserving MSSM.

\begin{figure}
\begin{center}
\includegraphics[width=8cm]{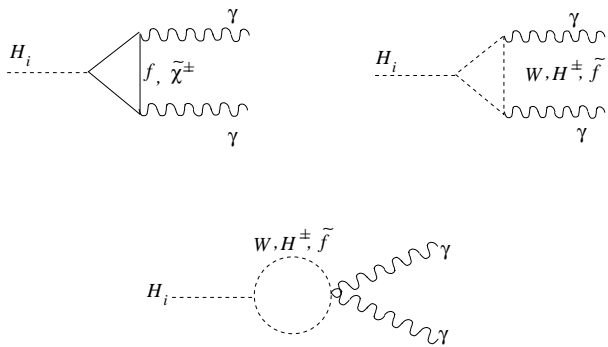}
\end{center}
\caption{Diagrams for Higgs decay into $\gamma\gamma$ pairs in the
  CP-violating MSSM:
$f\equiv t,\: b,\:\tau;\;\;\tilde f\equiv \tilde t_{1,2}, \tilde b_{1,2},
\tilde \tau_{1,2}$.}
\label{fig:HiggsPho}
\end{figure}

\begin{figure}
\includegraphics[width=8cm]{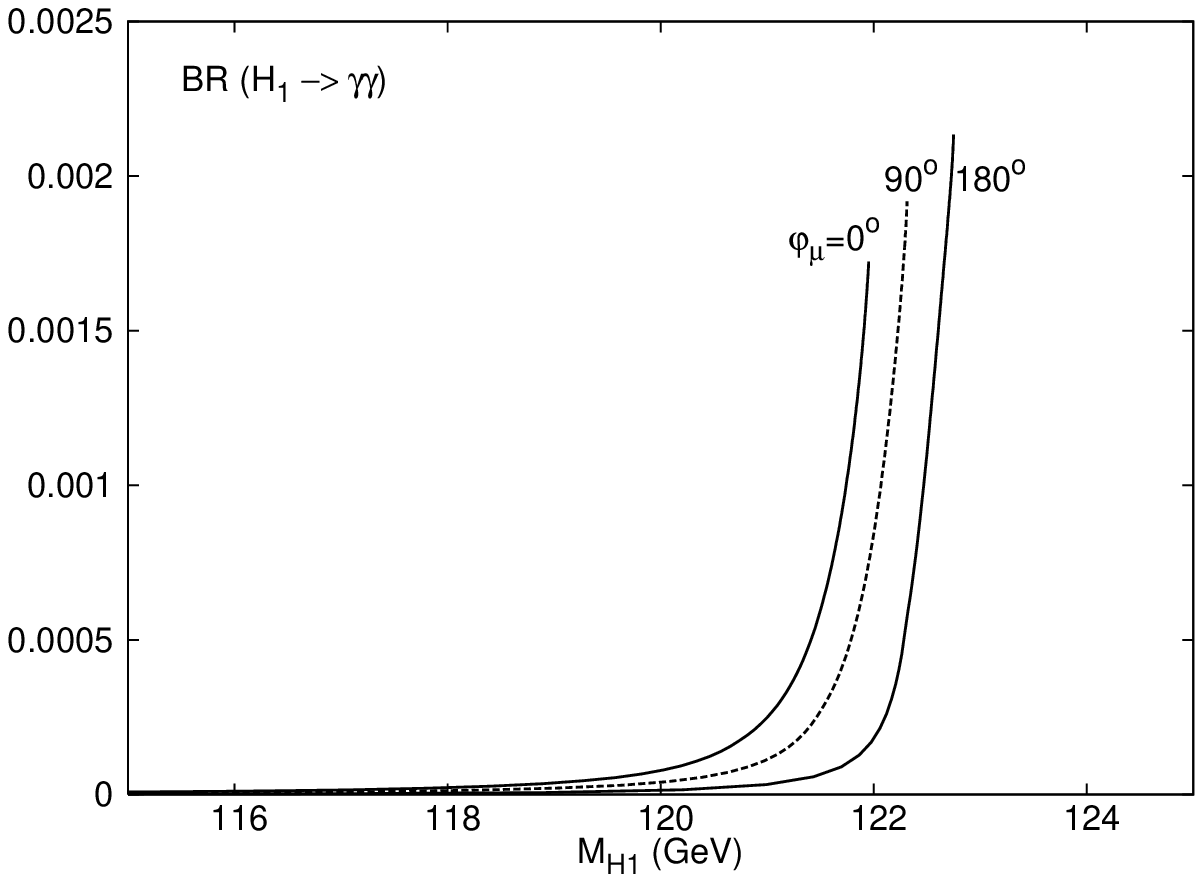}
\includegraphics[width=8cm]{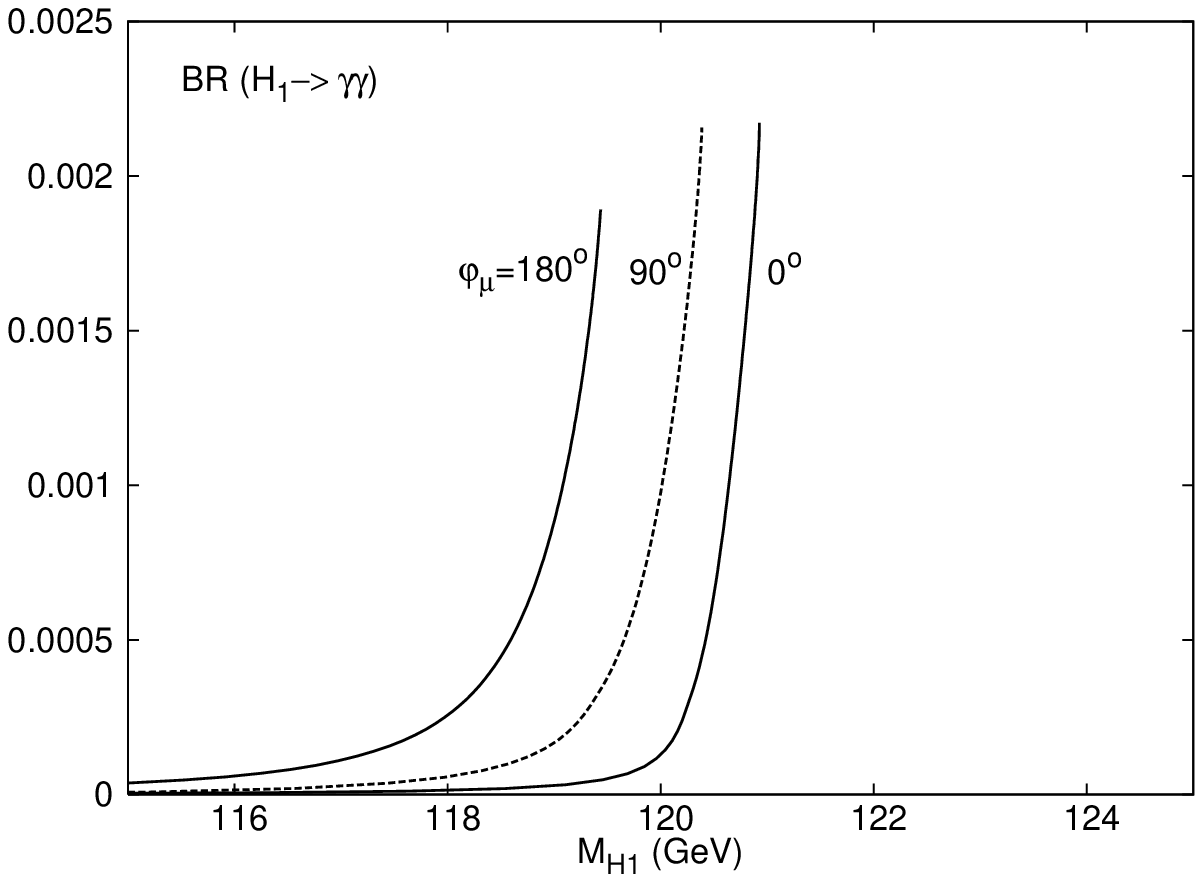}
\caption{BR($H_1\rightarrow \gamma\gamma$) plotted against $M_{H_1}$.
Parameters used are:
$\tan\beta=20,\;\; M_1=100\;{\rm GeV},\;\;M_2=M_3=1\;{\rm TeV},\;\;
M_{(\tilde Q_3, \tilde D_3, \tilde L_3, \tilde E_3)}=1\;{\rm TeV},\;\;
|\mu|=1\;{\rm TeV},\;\;
|A_f|= 1.5\;{\rm TeV}$. The upper plot is for
$M_{\tilde U_3}=1$ TeV while the lower one is for
$M_{\tilde U_3}=250$ GeV (the latter giving a rather light stop, ${m}_{\tilde t_1} = 200$ GeV).}
\label{fig:ssp}
\end{figure}

A Higgs boson in the MSSM decays at one-loop level into two photons
through loops of fermions, sfermions, $W^\pm$ bosons, charged Higgs
bosons and charginos, see Fig.~\ref{fig:HiggsPho}.
A random parameter space scan over about 100,000 parameter space points
to study the general
behaviour of the $\mathrm{BR}(H_1 \to \gamma\gamma)$ for non-zero
$\phi_\mu$ has revealed that about 50\% deviations are possible for
$M_{H_1}$ around 104 GeV for $\phi_\mu=100^\circ$.
In the considered mass range of 
90--130 GeV an average of 30\% deviation is found to occur.
Furthermore, this study of the average behaviour with and without a light
stop clearly establishes the strong impact
of a $\tilde{t}_1$ with a mass around 200 GeV on the deviations of the BR
\cite{Moretti:2007th}. Fig.~\ref{fig:ssp} illustrates this fact for a 
particular parameter set except for the stop mass and $\phi_\mu$, 
where $BR(H_1\rightarrow \gamma\gamma)$ is plotted against $M_{H_1}$ for 
different values of $\phi_\mu$ in the two cases of light and heavy stop.

\begin{figure}[t]
\includegraphics[width=18pc]{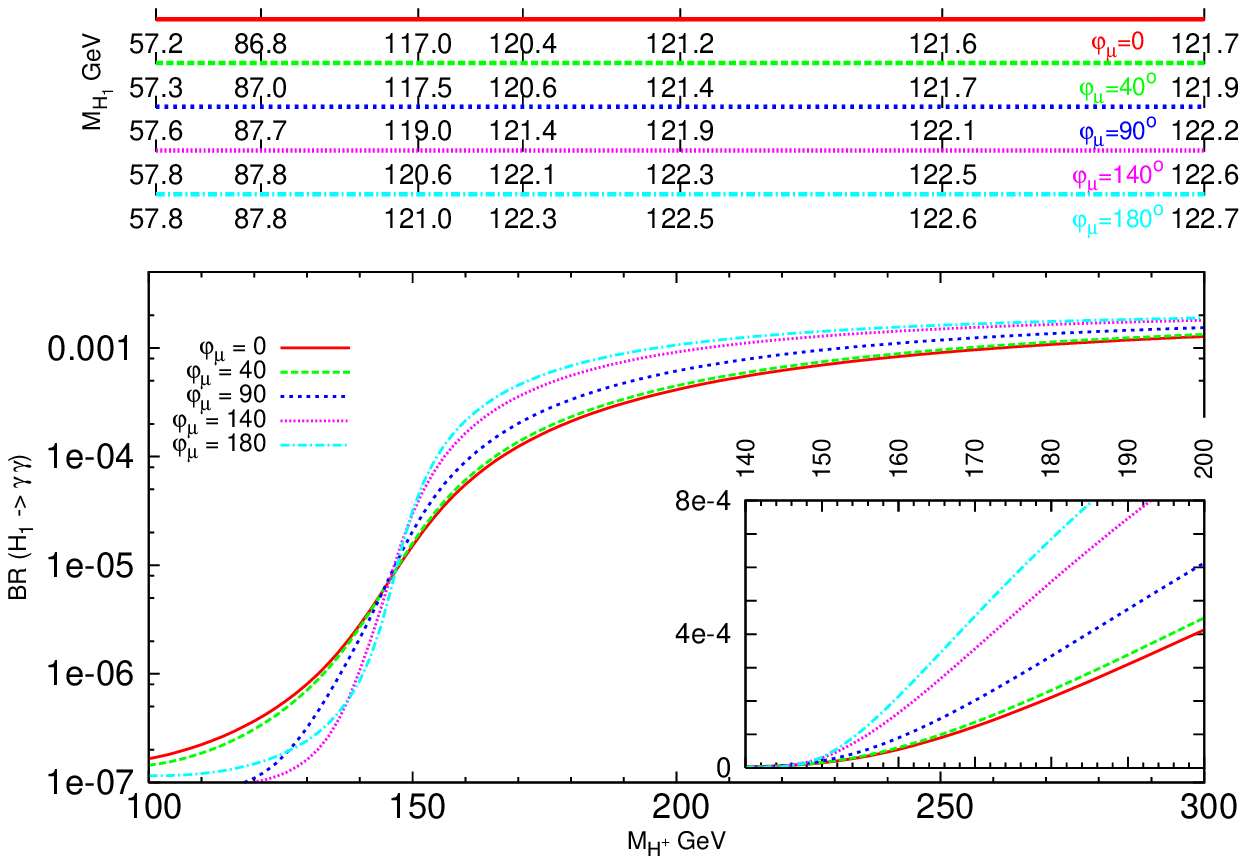}
\hspace{\fill}
\includegraphics[width=18pc]{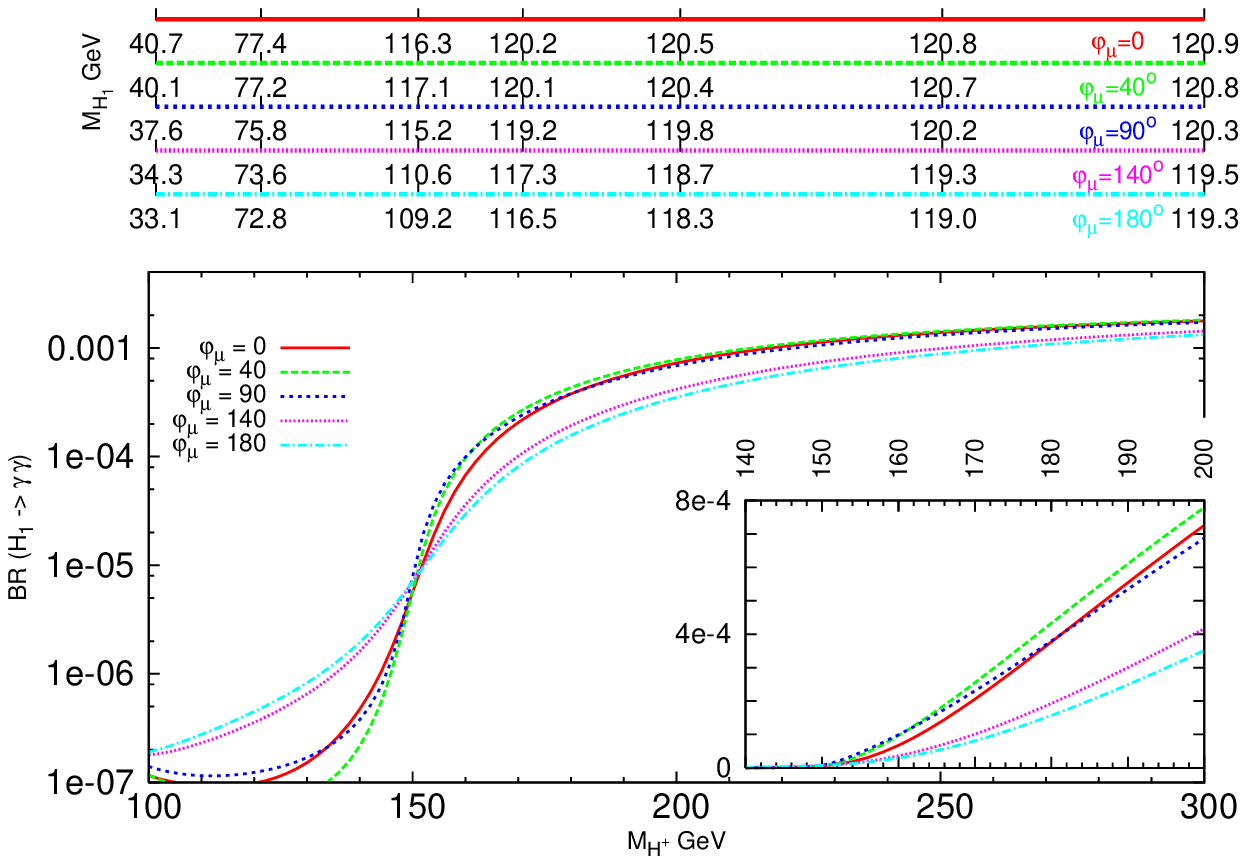}
\caption{\label{fig:BR}
BR of $H_1\rightarrow \gamma\gamma$
for $|A_f|=1.5$ TeV, $|\mu|=1$ TeV and $\tan\beta=20$.
Values of $M_{H_1}$ corresponding to representative points on the
$M_{H^+}$ axis
are indicated on the horizontal lines above separately for the values of
$\phi_\mu$ used.
The upper plot corresponds to the case with $M_{\tilde U_3}=1$~TeV
(no light SUSY particles),
while the lower plot corresponds to the case with
$M_{\tilde U_3}=250$~GeV (a light stop is present).}
\end{figure}

A detailed analysis at the matrix element level was undertaken in
 \cite{Hesselbach:2007en}, which consolidated the above observations. 
Since the mass of the Higgs particle itself changes by changing $\phi_\mu$ 
(and keeping
all other parameters the same), the difference in the BR read out from 
Fig.~\ref{fig:ssp} will have to be corrected for this change in $M_{H_1}$. 
(For the parameter set considered, 
the change in $M_{H_1}$ going from CP-conserving MSSM to CP-violating
MSSM, by changing the value of $\phi_\mu$, is within the typical experimental
uncertainty expected at LHC.)
In Fig.~\ref{fig:BR} we plot the BR($H_1\rightarrow \gamma\gamma$)
for five representative $\phi_\mu$ values between $0^\circ$ and $180^\circ$
as a function of $M_{H^+}$ for the two cases
$M_{\tilde U_3}=1$~TeV (all SUSY particles heavy) and
$M_{\tilde U_3}=250$ GeV (light $\tilde{t}_1$).
The respective values of $M_{H_1}$ are indicated separately on the horizontal
lines for each $\phi_\mu$ value.
The cross over point in the Higgs mass eigenstates at
$M_{H^+} \sim 150$ GeV is clearly visible. This corresponds to the sharp
rise of the BR at around $M_{H_1}\sim 120$ GeV in Fig.~\ref{fig:ssp}.
Below this point the BRs are very small and there is a strong $\phi_\mu$
dependence of $M_{H_1}$, hence our analysis is not relevant in this
parameter region.
Above $M_{H^+} \sim 150$~GeV and $M_{H_1} \gtrsim 115$~GeV,
the $\phi_\mu$ dependence of $M_{H_1}$ is within the expected
experimental uncertainty and
the BR is large enough to be important
for the LHC Higgs search.
In scenarios with heavy SUSY particles (upper plot) the BR increases with
increasing $\phi_\mu$ leading to a 50\% increase for
$\phi_\mu=90^\circ$ at $M_{H^+}\sim 200$ GeV.
This $\phi_\mu$ dependence is caused mainly by the $\phi_\mu$ dependence 
of the $H_1$ couplings to $W^\pm$ bosons and $t$ and $b$ quarks, which appear
in the loop-induced decay $H_1\rightarrow \gamma\gamma$.
When a light $\tilde{t}_1$ is present (lower plot)
the additional $\phi_\mu$
dependence in the stop sector causes a considerable change of the
$\phi_\mu$ dependence of the BR.
In fact, the BR increases again with increasing $\phi_\mu$ up to a maximum 
for some value of $\phi_\mu$ around $40^\circ$, beyond which, however, the
BR decreases to about 50\% at $\phi_\mu = 180^\circ$.

\begin{figure}[t]
\includegraphics[width=18pc]{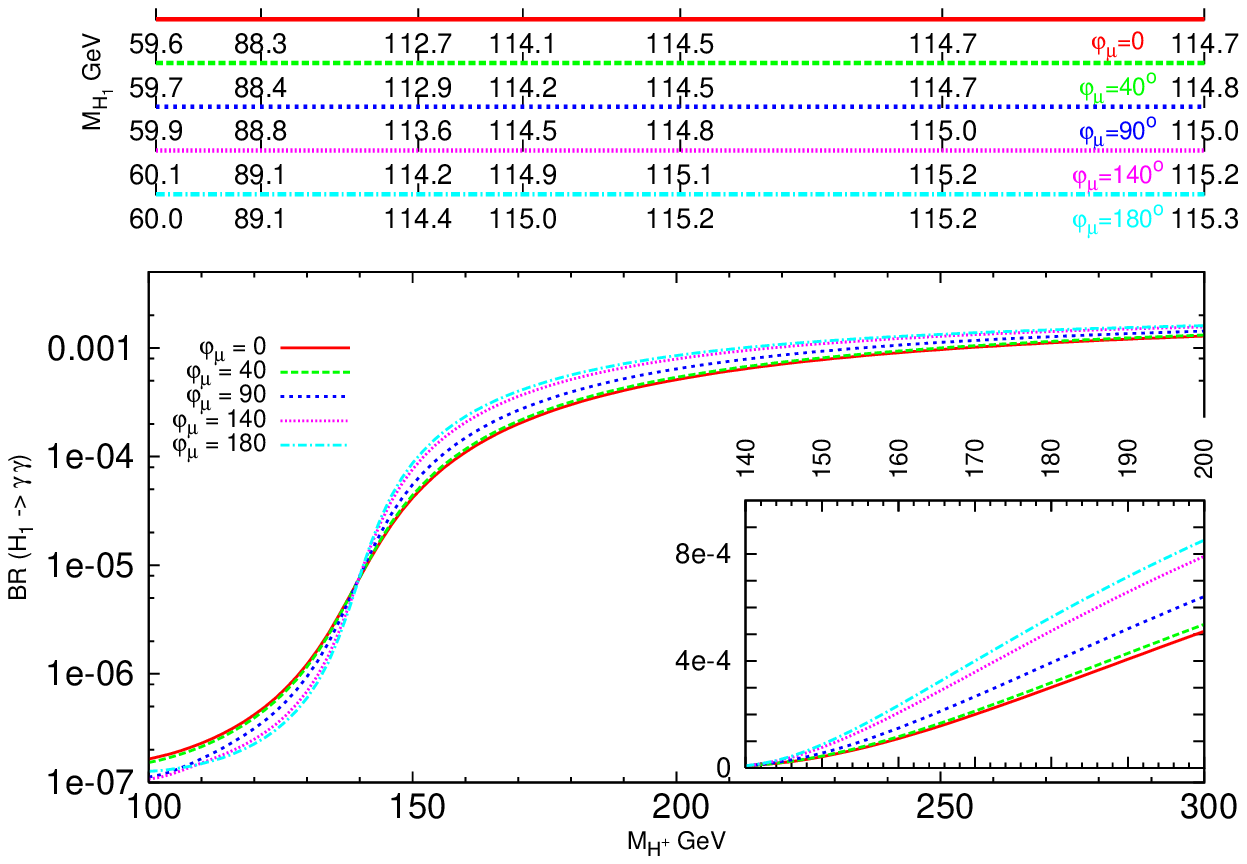}
\hspace{\fill}
\includegraphics[width=18pc]{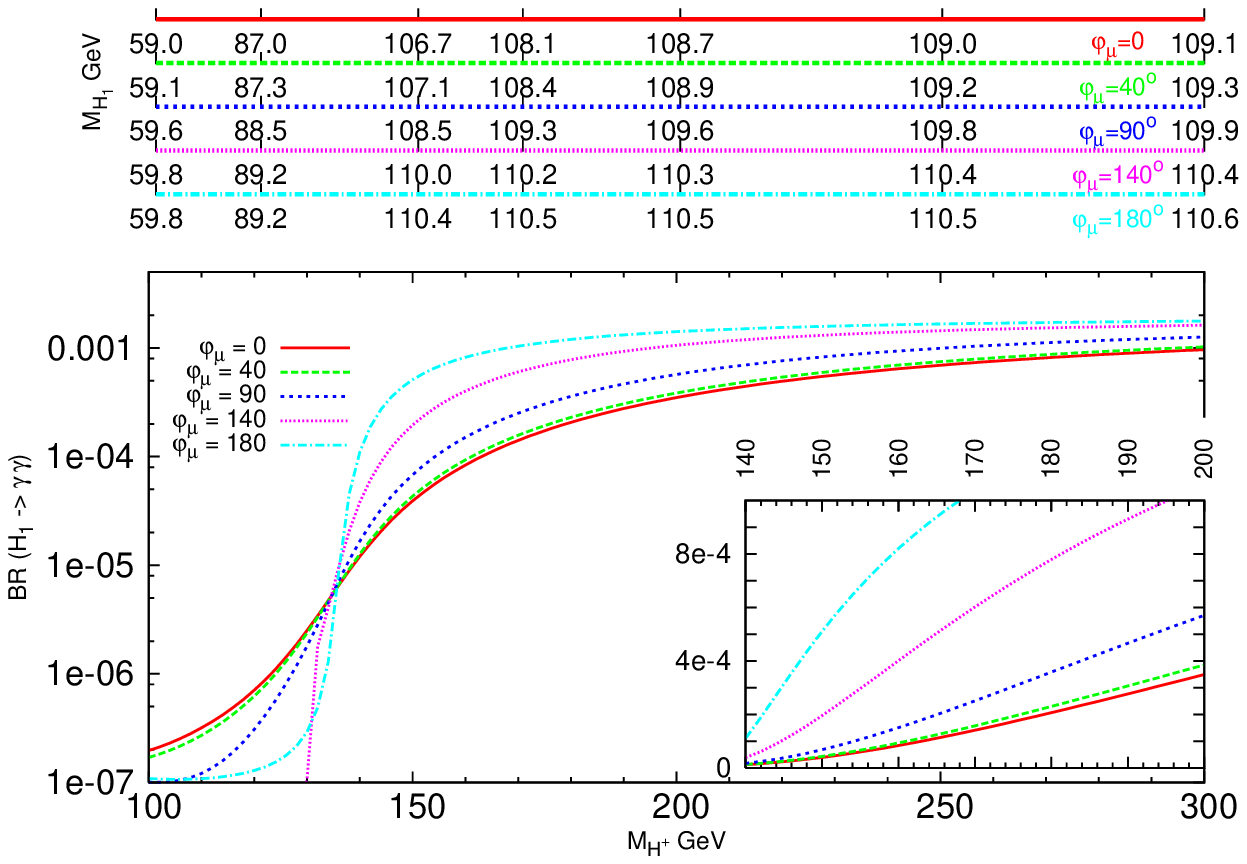}
\caption{\label{fig:BR_A500}
The same as Fig.~\ref{fig:BR} but with $|A_f|=0.5$ TeV, $|\mu|=1$ TeV
and $\tan\beta=20$.}
\end{figure}

\begin{figure}[t]
\includegraphics[width=18pc]{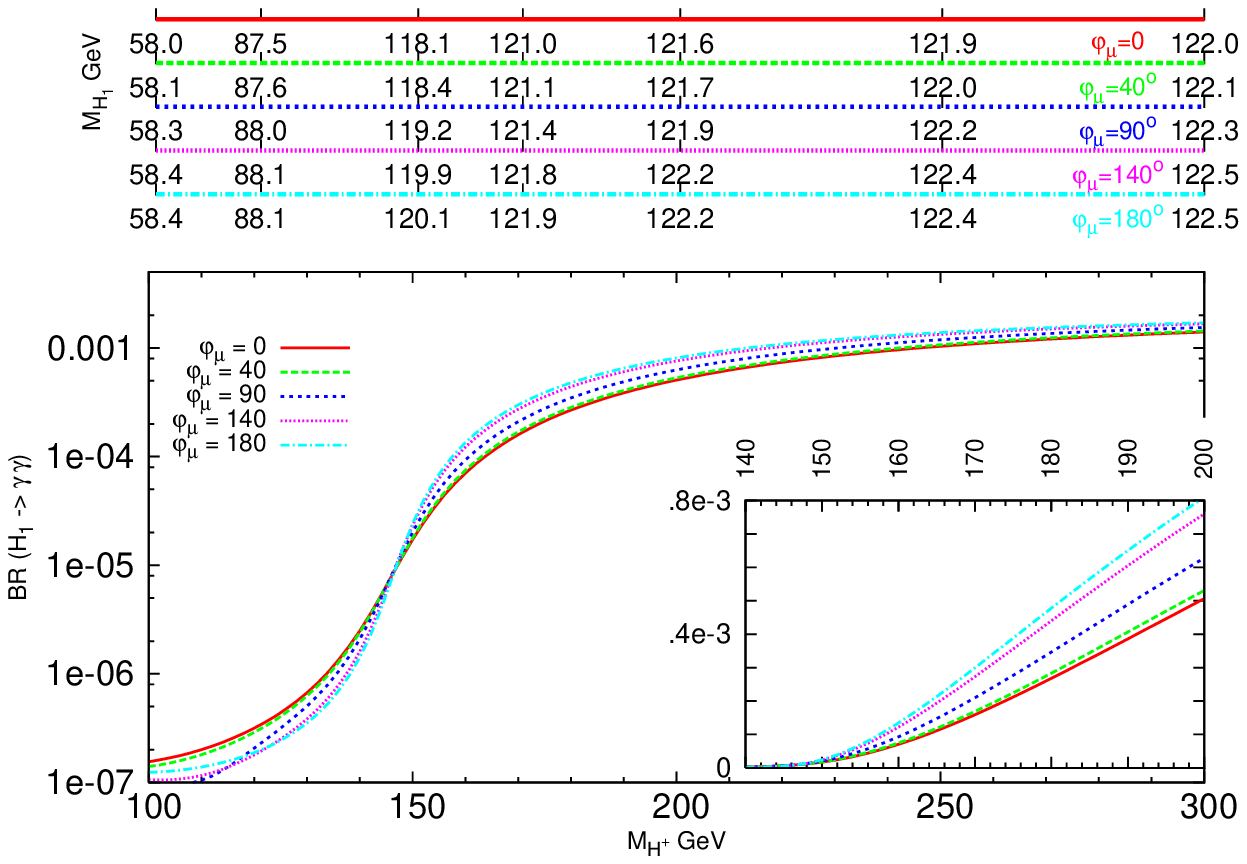}
\hspace{\fill}
\includegraphics[width=18pc]{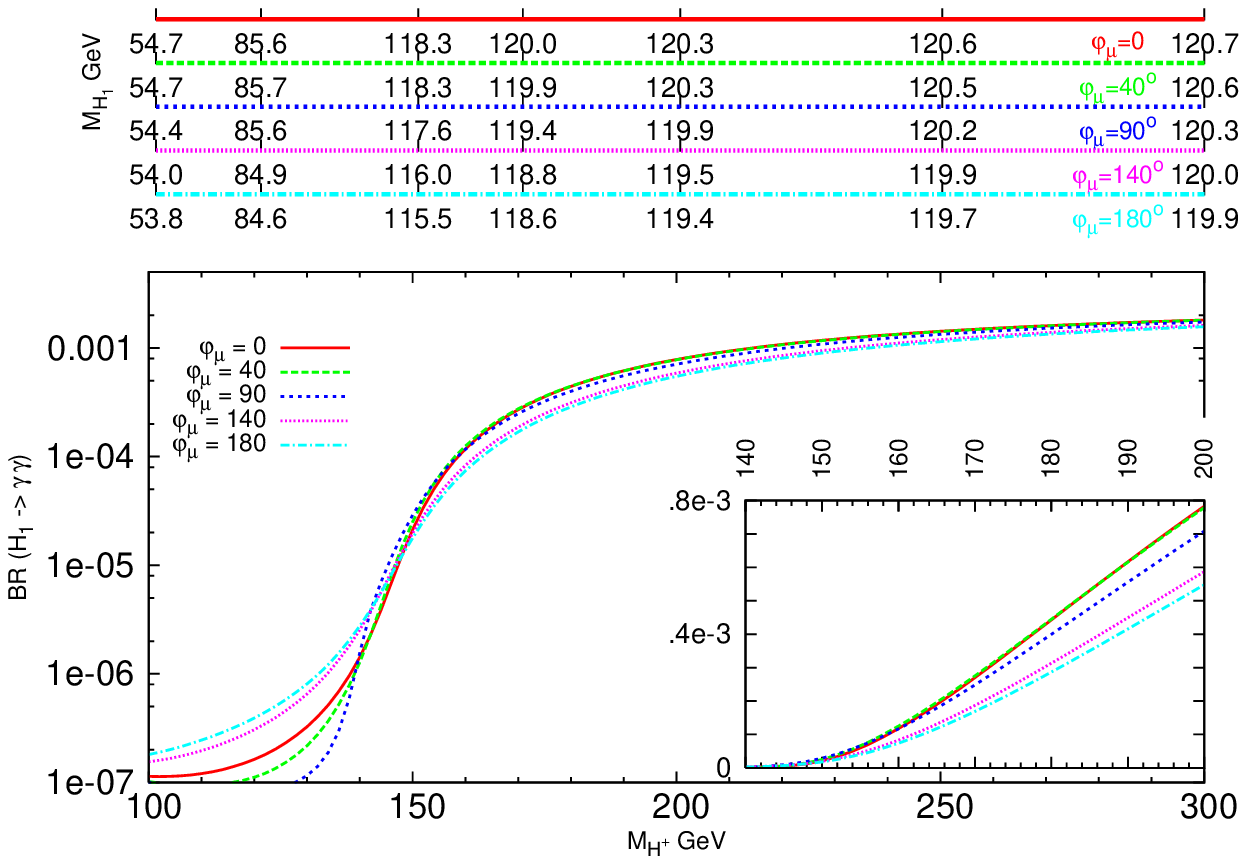}
\caption{\label{fig:BR_mu500}
The same as Fig.~\ref{fig:BR} but with $|A_f|=1.5$ TeV, $|\mu|=0.5$ TeV
and $\tan\beta=20$.}
\end{figure}

Our analysis with other relevant sparticles like sbottom and stau being light
shows that they do not play any major role in the $H_1\rightarrow
\gamma\gamma$ decay,
even for $\tan\beta$ values as large as 50. We have also taken care that
the changes in the masses of sparticles are not too large 
(i.e., again within expected experimental errors) when going from
$\phi_\mu=0$ to non-zero values of $\phi_\mu$ while keeping the other
parameters constant.
Concerning the dependence on other SUSY parameters, we have found
that a smaller value of $|A_f|$ considerably changes the $\phi_\mu$
dependence of the BR in scenarios with light a $\tilde{t}_1$
(see lower plot in Fig.~\ref{fig:BR_A500}) whereas
a smaller $|\mu|$ value leads generally to a smaller $\phi_\mu$
dependence (Fig.~\ref{fig:BR_mu500}). 

\section{Summary}

We have analysed the BR of the di-photon decay of the
lightest Higgs boson in the CP-violating MSSM with a complex $\mu$
parameter.
The presence of a light scalar top is found to influence the 
$\phi_\mu$ dependence
of the BR considerably while other sparticles have only
negligible effect.
In general, the BR may be increased or decreased for a non-zero
$\phi_\mu$ depending on the SUSY parameter point.
In scenarios where the relevant SUSY spectrum is already established
by the
LHC and consistent with both the CP-conserving and CP-violating scenario,
our analyses of $H_1\rightarrow \gamma\gamma$ will be able to distinguish
between these two cases.


\end{document}